\documentstyle[preprint,aps]{revtex}
\begin{document}
\draft
\title{Stars of WIMPs}
\author{F. Pisano\footnote{Supported by FAPESP} and
J. L. Tomazelli\footnote{Supported by CAPES}}
\address{Instituto de F\'\i sica Te\'orica \\
Universidade Estadual Paulista \\
Rua Pamplona, 145 \\
01405--000--S\~ao Paulo, SP \\
Brazil\\}
\maketitle
\begin{abstract}
A class of boson-fermion stars, whose spin-0 and spin-1/2 constituents
interact through a U(1) current-current term in the Lagrangian density, is
analysed. It is shown that it describes the low-energy behavior of a system
of weakly interacting massive particles (WIMPs) from the leptonic sector of
the minimal supersymmetric standard model. In this case the effective
coupling constant $\lambda$ is related to the Fermi constant $G_F$.
\end{abstract}
\narrowtext
\par
Along the last two decades there has been an increasing interplay between
gravity and elementary particle theories, in which the former has found a
field-theoretical support such as in the inflationary models~\cite{guth}.
These models predict that the density parameter of the Universe
$\Omega_{\rm total}=1$, in contrast with $\Omega_B \stackrel{<}{\sim}0.15$
for the baryonic density of the standard Big Bang nucleosynthesis. This
indicates the existence of nonbaryonic dark matter. Furthermore, there
is a direct evidence from astronomical observations of groups of galaxies and
clusters, whose dynamics indicate the presence of hidden
matter~\cite{stewart}.
Among the candidates for the dark matter are boson stars~\cite{ruff} and
composite structures of bosons and fermions~\cite{henr}, which might be formed
in the early Universe. However, none of the models considered up
to now succeeded in taking into
account any specific interaction among the constituents of such stellar
objects. Recently it was proposed a current-current type interaction between
those bosons and fermions~\cite{tomaz}. Here we reinterpret this interaction
in the context of the minimal supersymmetric standard model (MSSM) as an
effective interaction corresponding to the exchange of a massive gauge boson
and determine the strength of the coupling. In this case the fermion and
scalar particles are massive charged leptons together with their respective
supersymmetric partners, a particular set of weakly
interacting massive particles (WIMPs) which are the prime condidates to the
above mentioned nonbaryonic dark matter.
\par
Let us start with the Lagrangian density for pure gravity and the free
complex scalar field without self-coupling,
\begin{equation}
{\cal L}= \frac{R}{16\pi G}
+\partial_{\mu}\Phi ^*\partial ^{\mu}\Phi
+m_s^{2}\Phi^*\Phi \,\,,
\end{equation}
where
\begin{equation}
\Phi(r,\tau)=\phi(r)e^{-i\omega \tau}
\end{equation}
and $\phi$ is a complex scalar field.

The background fermion field is described by a perfect fluid~\cite{chandra}
with
energy density $\rho$ and pressure $p$:

\begin{eqnarray}
\rho &=&K (\sinh t -t) \,\,,                             \nonumber \\
                                                         \\
   p &=& \frac{K}{3}(\sinh t -8\sinh\frac{t}{2}+3t)\,\,, \nonumber
\end{eqnarray}
where $t$ is a parameter, $K=m^{4}_{f}/32\pi^{2}$, and $m_{f}$ is the
fermion mass.

The corresponding energy-momentum tensor for the free fields reads as
\begin{equation}
T_{\mu \nu}^{(0)}=T_{\mu \nu}^{s}+T_{\mu \nu}^{f} \,\,,
\end{equation}
with
\begin{eqnarray}
T_{\mu \nu}^{s} &=& \partial_{\nu}\Phi ^*\partial_{\mu}\Phi +
\partial_{\mu}\Phi ^*\partial_{\nu}\Phi
-g_{\mu\nu}(\partial_{\lambda}\Phi^* \partial ^{\lambda}\Phi
+m_s^2{\Phi}^*\Phi ) \,\,,  \\
T_{\mu \nu}^{f} &=& (\rho+p)u_{\mu}u_{\nu}-pg_{\mu \nu}  \,\,,
\end{eqnarray}
where indices $s$ and $f$ label the scalar and fermion fields,
respectively.

The complete Lagrangian density includes the contact interaction term
\begin{equation}
{\cal L}^{\rm int}=-\lambda j_{\mu}(\psi)J^{\mu}(\Phi) \,\,,
\end{equation}
where
\begin{eqnarray}
j_{\mu}(\psi) &=& \overline{\psi}{\gamma}_{\mu}\psi \,\,, \\
J^{\mu}(\Phi) &=& i({\Phi}^*\partial^{\mu}\Phi -
\Phi \partial^{\mu}{\Phi}^*) \,\,.
\end{eqnarray}

It is invariant under global U(1) gauge transformations and contributes
to the energy-momentum tensor with
\begin{equation}
T_{\mu \nu}^{\rm int}=-\lambda( J_{\mu}j_{\nu}-g_{\mu\nu}
j_{\alpha}J^{\alpha}) \,\,.
\end{equation}

In the region of low frequencies the fermion-current can be replaced by
\begin{equation}
j_{\mu}(\psi)=\overline{\psi}\Gamma_{\mu}\psi \,\,,
\end{equation}
where
\begin{equation}
\Gamma_{0} \equiv u_0 \,\,\,\,\,,\,\,\,\,\,\Gamma_i \equiv iu_i \,\,.
\end{equation}
The four-vector $u_{\mu}=(u_0,u_r,u_{\theta},u_{\varphi})$ is the
four-velocity of the fermion fluid in spherical coordinates. Our metric
convention is ($+,-,-,-$) otherwise specified.
%
\par
Let us pick out from the interaction Lagrangian density for the broken
supersymmetric SU(2)$\otimes$U(1) theory~\cite{kane} those terms that
contribute to the scattering of massive leptons and scalar-leptons, which
can be identified with the effective interaction Lagrangian density (7) in
the low-energy limit:
\begin{equation}
{\cal L}^{\rm int}_{l\tilde{l}\rightarrow l\tilde{l}}={\cal L}_{llV}
+{\cal L}_{\tilde{l}\tilde{l}V}\,\,,
\end{equation}
with
\begin{eqnarray}
{\cal L}_{llV}                     &=& -\frac{g}{2\cos\,\theta_W}\overline{l}
\gamma_{\mu}(g_V-g_A\gamma_5)l\,Z^{\mu}+e\overline{l}\gamma_{\mu}l\,
A^{\mu}\,\,,\\
{\cal L}_{\tilde{l}\tilde{l}V}     &=&
-i\frac{g}{2\cos\,\theta_W}\tilde{l}^{* T}\stackrel{\leftrightarrow}
{\partial}_{\mu}(\tilde{g}_V-\tilde{g}_A\Gamma_5)\tilde{l}\,Z^{\mu}+ie
\tilde{l}^{* T}\stackrel{\leftrightarrow}{\partial}_{\mu}
\tilde{l}\,A^{\mu}\,\,,
\end{eqnarray}
where we have defined
\begin{equation}
\Gamma_5 \equiv
\left(
\begin{array}{cc}
-1 & 0 \\
 0 & 1
\end{array}
\right) \,,\,\,\,\,\,
\tilde{l} \equiv \left(
\begin{array}{c}
\tilde{l}_L \\
\tilde{l}_R
\end{array}
\right)
\end{equation}
and
\begin{equation}
g_V=\tilde{g}_V \equiv -\frac{1}{2}+2\sin^2\,\theta_W\,\,\,\,{\rm and}
\,\,\,\, g_A=\tilde{g}_A \equiv -\frac{1}{2}\,\,.
\end{equation}

These terms give rise to the interaction between a fermion- and a
scalar-current mediated by a vector gauge boson. Since (7) is a contact
interaction, the exchanged virtual particle must be massive. Thus, neither
the last term in (14) nor that in (15) must be taken into account, since they
produce a long-range interaction. The scalar neutral-currents that comprise
the interaction Lagrangian density (15) were written in a suggestive fashion
in order to stress the correspondence with the respective fermion
counterparts in (14), where $\tilde{l}_L$ and $\tilde{l}_R$ stand for the
supersymmetric partners of the ordinary charged lepton states
$l=\{e,\,\mu,\,\tau\}$ projected in the chiral basis.

For the moment, we consider processes involving electrons and
scalar-electrons. In the MSSM such scalar particles must correspond
to fermions in a definite chirality state. The amplitudes for the
$e^{-}\tilde{e}_R^{-}\rightarrow e^{-}\tilde{e}_R^{-}$ and
$e^{-}\tilde{e}_L^{-}\rightarrow e^{-}\tilde{e}_L^{-}$ processes in momentum
space are
\begin{eqnarray}
{\cal M}^R &=& -ig^2\displaystyle{\frac{\tan^2\theta_W}{2}}
\overline{u}(\vec{p}_1{}^{\prime})(p_2\!\!\!\!\!\slash+p'_2\!\!\!\!\!\slash
\,\,)(g_V-g_A\gamma_5)\,u(\vec{p}_1)\frac{1}{k^2-M_Z^2}\,, \\
{\cal M}^L &=& -ig^2\displaystyle{\frac{1+2\sin^2\theta_W}{4\cos^2\theta_W}}
\overline{u}(\vec{p}_1{}^{\prime})(p_2\!\!\!\!\!\slash+p'_2\!\!\!\!\!\slash
\,\,)(g_V-g_A\gamma_5)\,u(\vec{p}_1)\frac{1}{k^2-M_Z^2},
\end{eqnarray}
where $k^2=(p'_1-p_1)^2$ and we have written the massive gauge boson
propagator in the Feynman gauge.
%
\par
At the infrared, where spin effects become negligible, we can safely
disregard the piece of (18) and (19) which contains the axial-current. In
fact, it follows from the Dirac equation
$$(\slash\!\!\!p-m)u_r(\vec{p})=0\,;\,\,r=1,2$$
for positive-energy four-spinors and relation
$$\sigma^{ij}=-\gamma^0\gamma^5\gamma^k \,;\,\,k=1,2,3\,\,\,\,\,
{\rm in\,\, cyclic\,\, order}$$
that
\begin{equation}
\overline{u}_s(\vec{p}\,{}^{\prime})\gamma^{\mu}\gamma^5u_r(\vec{p})=
\overline{u}_s(\vec{p}\,{}^{\prime})\gamma^{\mu}\left[\frac{p_0-m\gamma^0}
{|\vec{p}|}\right](-1)^{r+1}u_r(\vec{p})\,\,,
\end{equation}
where $u_r(\vec{p})$ are eigenstates of the helicity operator
$$\sigma_{\vec{p}}=\frac{\vec{\sigma}.\vec{p}}{|\vec{p}|}\,\,.$$
Making the replacement $\gamma^{\mu}\rightarrow \Gamma^{\mu}$, where the
$\Gamma$'s were defined in (12), we immediately see that the axial-current
vanishes.

Hereafter we shall consider only the vector-current amplitudes, which for
$k^2 \ll M_Z^2$ reduce to
\begin{eqnarray}
{\cal M}^R &=& i\frac{g^2}{4M_Z^2}\tan^2\theta_W(-1+4\sin^2\theta_W)
\overline{u}(\vec{p}_1{}^{\prime})(p_2\!\!\!\!\!\slash+p'_2\!\!\!\!\!
\slash\,\,)\,u(\vec{p}_1)\,, \\
{\cal M}^L &=& i\frac{g^2}{4M_Z^2}\displaystyle{\frac{(1+2\sin^2\theta_W)
(-1+4\sin^2\theta_W)}
{2\cos^2\theta_W}}
\overline{u}(\vec{p}_1{}^{\prime})(p_2\!\!\!\!\!\slash+p'_2\!\!\!\!\!
\slash\,\,)\,u(\vec{p}_1)\,.
\end{eqnarray}

The ratio between the total cross-sections constructed from amplitudes (21)
and (22) is given by
\begin{equation}
\frac{\sigma_R}{\sigma_L}=\left(\frac{2x}{1+2x}\right)^2 \,\,,\\
\end{equation}
with $x \equiv \sin^2\theta_W$. Since the value of $x$ decreases
with energy, the first process is supressed in the infrared limit, where
we have
$$\frac{\sigma_R}{\sigma_L} \sim 4x^2-16x^3+48x^4-\dots\,,\,\,\,x < 1\,\,.$$

If we compare the amplitude which stems from the effective interaction (7)
with (22), we extract the value
\begin{eqnarray}
\lambda &=& \frac{g^2}{4M_Z^2}\frac{(1+2\sin^2\theta_W)
(1-4\sin^2\theta_W)}{2\cos^2\theta_W} \nonumber \\
        &=& \frac{G_F}{4\sqrt{2}}(1+2\sin^2\theta_W)(1-4\sin^2\theta_W)\,\,,
\end{eqnarray}
where $G_F$ is the Fermi constant.

The effective coupling constant $\lambda$ can be expressed in terms of the
adimensional coupling $\overline{\alpha}$ that appears in the Einstein
equations for the metric coefficients and the Klein-Gordon equation for the
radial component of the scalar field~\cite{tomaz}
\begin{equation}
\lambda=\frac{m_s\overline{\alpha}}{2\overline{n}_f}\,\,,
\end{equation}
where $\overline{n}_f$ is the average fermion density. From (24) and (25)
we have
\begin{equation}
\overline{\alpha}=\frac{G_F}{2\sqrt{2}}\frac{\overline{n}_f}{m_s}
(1+2\sin^2\theta_W)(1-4\sin^2\theta_W)\,\,.
\end{equation}
All the scalar-leptons have essentialy the same lower bound for their masses,
$m_s \sim 45$ GeV~\cite{pdg}, so that $\overline{\alpha}$ is independent of
the scalar-lepton flavor. For a background of a degenerate electron
gas~\cite{zeld}, $\overline{n}_f$ is of order $10^{30}$ ${\rm cm}^{-3}$.
Substituting the values for $G_F$, $x$ and $m_s$, we obtain
\begin{equation}
\overline{\alpha} \stackrel{<}{\sim} 10^{-19}\,\,
\end{equation}
and therefore we do not reach the critical value for the second-order phase
transition reported in~\cite{tomaz}, since the electron density is not high
enough to
create bound states of electrons and scalar-electrons.
%
\par
Summing up we have investigated gravity in the presence of scalar fields in
interaction
with a background fermion fluid, in the framework of a fundamental particle
theory, namely the MSSM in which the Bose-Fermi system can be interpreted as
a self-gravitating system of WIMPs, the strongest candidates to the dark
matter. This enabled us to attribute a precise meaning to the arbitrary
coupling constant $\lambda$ of the effective theory, which results to be
proportional to the Fermi constant $G_F$ and the electroweak mixing parameter
$\sin^2\theta_W$. For a degenerate Fermi gas the value of the adimensional
coupling $\overline{\alpha}$ is bellow the critical value for Bose-Einstein
condensation.


\begin{references}
\bibitem{guth} A.\ H.\ Guth, Phys.\ Rev.\ {\bf D 23}, 347 (1981);
             A.\ D.\ Linde, Phy.\ Lett.\ {\bf B 108}, 347 (1982);
	     A.\ Albrecht and P.\ J.\ Steinhardt, Phy.\ Rev.\ Lett.\
	     {\bf 48}, 1220 (1982).
\bibitem{stewart} G.\ C.\ Stewart, C.\ R.\ Canizares, A.\ C.\ Fabian
and P.\ E.\ J.\ Nilsen, Astrophys.\ J.\ {\bf 278}, 53 (1984).
\bibitem{ruff} R.\ Ruffini and S.\ Bonazzola, Phys.\ Rev.\ {\bf 187}, 1767
        (1969).
\bibitem{henr} A.\ B.\ Henriques, A.\ R.\ Liddle and R.\ G.\ Moorhouse,
Phys.\ Lett.\ {\bf B 233}, 99 (1989).
\bibitem{tomaz} C.\ M.\ G.\ de Sousa and J.\ L.\ Tomazelli,
                preprint gr-qc/9507043.
\bibitem{chandra} S. Chandrasekhar, Astrophys.\ J.\ {\bf 140}, 417 (1964).
\bibitem{kane} H.\ E.\ Haber and G.\ L.\ Kane, Phys.\ Rep.\ {\bf C 117}, 75
               (1985).
\bibitem{pdg} L.\ Montanet {\sl et al.} (Particle Data Group), Phys. Rev.
              {\bf D 50}, 1173 (1994).
\bibitem{zeld} Ya. B. Zeldovich and I. D. Novikov, {\em Relativistic
             Astrophysics}, Vol.{\bf 1}, The University of Chicago Press
	     (1971).
\end{references}
\end{document}